\documentclass{emulateapj}
\usepackage{bm}
\usepackage{chngpage}
\usepackage{graphicx}
\usepackage[caption=false]{subfig}
\usepackage{mathrsfs}
\usepackage{setspace}

\def\apjl{ApJL}

\shorttitle{FRB 200428}
\shortauthors{Geng et al.}

\begin{document}

\title{FRB 200428: an Impact between an Asteroid and a Magnetar}

\author{Jin-Jun Geng\altaffilmark{1,2}, Bing Li\altaffilmark{3,4}, Long-Biao Li\altaffilmark{5,6}, Shao-Lin Xiong\altaffilmark{3}, Rolf Kuiper\altaffilmark{2} and Yong-Feng Huang\altaffilmark{1,6}}

\altaffiltext{1}{School of Astronomy and Space Science, Nanjing University, Nanjing 210023, China; gengjinjun@nju.edu.cn}
\altaffiltext{2}{Institute of Astronomy and Astrophysics, University of T\"ubingen, Auf der Morgenstelle 10, D-72076,T\"ubingen, Germany}
\altaffiltext{3}{Key Laboratory of Particle Astrophysics, Institute of High Energy Physics, Chinese Academy of Sciences, Beijing 100049, China; libing@ihep.ac.cn}
\altaffiltext{4}{Particle Astrophysics Division, Institute of High Energy Physics, Chinese Academy of Sciences, Beijing 100049, China}
\altaffiltext{5}{School of Mathematics and Physics, Hebei University of Engineering, Handan 056005, China}
\altaffiltext{6}{Key Laboratory of Modern Astronomy and Astrophysics (Nanjing University), Ministry of Education, Nanjing 210023, China}

\begin{abstract}
A fast radio burst (FRB) was recently detected to be associated with a hard X-ray burst from the Galactic magnetar SGR 1935+2154.
Scenarios involving magnetars for FRBs are hence highly favored.
In this work, we suggest that the impact between an asteroid and a magnetar could explain such a detection.
According to our calculations, an asteroid of mass $10^{20}$~g will be disrupted at a distance of $7 \times 10^9$~cm when approaching the magnetar.
The accreted material will flow along the magnetic field lines from the Alfv\'en radius $\sim 10^7$~cm.
After falling onto the magnetar's surface, an instant accretion column will be formed, 
producing a Comptonized X-ray burst and an FRB in the magnetosphere. 
We show that all the observational features of FRB 200428 could be interpreted self-consistently in this scenario.
We predict quasi-periodic oscillations in this specific X-ray burst, which can serve as an independent observational test.

\end{abstract}


\section{Introduction}

The physical origin of fast radio bursts (FRBs) has long been remaining mysterious.
Their high dispersion measures indicate an extragalactic origin in the past years,
and the corresponding isotropic energy released is then $10^{39-40}$ erg.
Many catastrophic models like compact object mergers/interactions, the collapse of compact objects,
or giant flares/pulses in magnetars, etc.~are proposed to explain FRBs 
(see \citealt{Platts19} for a living review of these models).
The observational data of FRBs are fastly growing thanks to the efforts of many telescopes (e.g., Canadian Hydrogen Intensity Mapping Experiment, CHIME).
The first FRB was discovered by \cite{Lorimer07}.
As the number of FRBs increased, it was found that some of them 
are individual events, i.e., do not repeat within a monitor period. 
On the other hand, some sources are repeating since discovery. 
The repeating FRB 121102 has been bursting in a seemingly irregular/cluster~\footnote{\cite{Rajwade20}
reported a detection of tentative periodic behaviour of FRB 121102 over the span of
five years of data, which invokes further follow-up observations to affirm.} pattern for over 7 years~\citep{Spitler16}.
In addition, the repeating FRB 180916.J0158+65 shows $\sim 16$-day periodic activity~\citep{CHIME20a},
which strongly implies the source is modulated by the orbital motion of a binary system.
It is still uncertain whether repeating or ``non-repeating'' is due to observational bias,
e.g, we are not in the most active window for the individual FRB, or different astrophysical origin sources~\citep{Li19,Palaniswamy18}. 

Recently, a fast radio burst has been reported to be 
spatially coincident with the galactic Soft Gamma-ray Repeater (SGR) 1935+2154~\citep{CHIME20b,Lin20b,Bochenek20},
also associated with a hard X-ray burst detected by several instruments including 
INTEGRAL \citep{Mereghetti20}, {\it Insight}-HXMT \citep{Li20}, AGILE \citep{Tavani20}, 
NICER \citep{Pearlman20}, and Konus-Wind \citep{Ridnaia20}.
Although this FRB (FRB 200428 hereafter) is $\sim$ 25 times less energetic than the weakest FRB previously detected 
from cosmological FRB sources, the association of FRB 200428 with SGR 1935+2154 
supports the idea that the magnetar scenario should work for at least some FRBs.
However, detailed analyses indicate that a single magnetar model is very difficult to account for 
the observational properties of all FRBs \citep{Margalit20}. 
Therefore, several models may be expected for FRBs, which invokes authors to refine their specific models
to address corresponding characteristics and predictions (e.g., \citealt{Lu20}).

It has been proposed that small solid bodies such as asteroids or comets can occasionally impact 
neutron stars (NSs), which may trigger the energy release in the NS magnetosphere \citep{Colgate81,Huang14}
and hence FRBs~\citep{Geng15}. 
Within this scenario, the predicted X-ray radiation from heated crust after the collision 
is generally too faint to be detected at the cosmological distance.
\cite{Dai16} studied the electron acceleration mechanism during the collision process in great detail,
and suggested that the repeatability of FRB 121102 can be explained as due to multiple asteroid collision events.
Very recently, \cite{Dai20} argued that the observational features of FRB 200428 and its associated X-ray burst 
can be well explained by the magnetar-asteroid collision model. Especially, he suggested that the two pulses of 
FRB 200428 are produced when two major iron-nickel fragments of the disrupted asteroid cross the magnetic lines around the Alfv\'en radius.
Here, we further study the interaction between the asteroids and the NSs
by focusing on the formation of the instant accretion column and the consequent emissions.
The association of galactic FRB 200428 with a $\sim 10$~ms X-ray burst motivates us to 
test our scenario, especially in the framework of the magnetar, which has much stronger magnetic fields.

The structure of this article is as follows. We describe the interaction processes
in Section 2. In Section 3, we calculate the resulting emission from the interaction.
Finally, in Section 4, we discuss the implication of our model and summarize our conclusions.

\section{Magnetar-Asteroid Interaction}

SGR 1935+2154 is a magnetar with rotational period $P = 3.24$~s~\citep{Stamatikos14},
a surface dipolar magnetic field $B_{\rm NS} \simeq 2.2 \times 10^{14}$~G derived from
the spin-down rate $\dot{P} \simeq 1.43 \times 10^{-11}$~s~s$^{-1}$~\citep{Israel16}.
It is associated with the supernova remnant (SNR) G57.2+0.8~\citep{Gaensler14},
of which the distance is roughly within a range of 6-15 kpc \citep{Sun11,Pavlovic13,Kothes18,Zhong20,Zhou20}.
In this work, we adopt $D_L = 10$~kpc for SGR 1935+2154, and the uncertainty of $D_L$ will not affect our results significantly. 
This magnetar has exhibited multiple episodes of outbursts since discovery~\citep{Lin20a}.

When an asteroid approches a magnetar of mass $M_{\rm NS}$, it will be tidally disrupted due to gravity. 
For an Fe–Ni asteroid with a mass $m_{\rm a}$, density $\rho_{\rm a}$, radius $r_{\rm a}$, and shear strength $s$, 
its breakup radius is~\citep{Colgate81}
\begin{eqnarray}
R_b &=& (\rho_{\rm a} r_{\rm a}^2 M_{\rm NS} G / s)^{1/3} \\ \nonumber
    &=& 6.8 \times 10^9 m_{\mathrm{a},20}^{2/9} \rho_{\mathrm{a},0.9}^{1/9} s_{10}^{-1/3} M_{\mathrm{NS},1.4 M_{\odot}}^{1/3} \mathrm{cm},
\end{eqnarray}
where $G$ is the gravitational constant. 
The convention $Q_x = Q/10^x$ in cgs units is adopted hereafter.
The light cylinder of the magnetar is
\begin{equation}
R_{\rm LC} = c/\Omega = 1.5 \times 10^{10} \mathrm{cm},
\end{equation}
where $\Omega = 2 \pi/P$ is the angular velocity of the magnetar.
Since $R_b \le R_{\rm LC}$, the disruption of the asteroid starts roughly within the magnetosphere.
As the asteroid material gets much closer to the magnetar, magnetic field begins to disturb the inflow
at the so-called Alfv{\'e}n radius $R_{\rm A}$.
The equilibrium of kinetic and magnetic energy 
\begin{equation}
\rho_{\rm a} \frac{v_{\rm ff}(R_{\rm A})^2}{2} = \frac{B^2(R_{\rm A})}{8 \pi}
\end{equation}
gives the Alfv{\'e}n radius 
\begin{equation}
R_{\mathrm A} = 1.1 \times 10^7 B_{\mathrm{NS},14}^{2/5} R_{\mathrm{NS},6}^{6/5} \rho_{\mathrm{a},0.9}^{-1/5} M_{\mathrm{NS}, 1.4 M_{\odot}}^{-1/5} \mathrm{cm},
\end{equation}
where $v_{\rm ff}$ is the free-fall velocity, $R_{\rm NS}$ is the radius of the magnetar,
and a dipole magnetic field $B \propto R^{-3}$ is adopted.
Within the Alfv\'en radius we expect that the matter will flow along field lines.
The difference of arrival time at the magnetar surface for the leading and lagging fragments is
$\sim 10~m_{\mathrm{a},20}^{4/9} s_{10}^{-1/6} \rho_{\mathrm{a},0.9}^{-5/18} M_{\mathrm{NS},1.4 M_{\odot}}^{-1/3}$~ms~\citep{Geng15},
which is less than the active duration of the radio/X-ray burst. 

It should be noted that the magnetar-asteroid interaction above requires a relatively low magnetar surface temperature
to avoid the evaporation of asteroid (see details in \citealt{Cordes08}) before its breakup.
The NS surface temperature is likely to be $10^5-10^6$ K in the literature but remains very uncertain.
Considering the age of the supernova remnant G57.2+0.8 hosting the magnetar SGR 1935+2154 is $\ge 2 \times 10^4$ yr~\citep{Zhou20},
the magnetar surface temperature could be $\le 10^{5.4}$ K for a magnetar mass of $\ge 1.5 M_{\odot}$
because cooling is much more efficient for high-mass NSs~\citep{Harding06}.
Such a surface temperature could ensure that the asteroid temperature in thermal equilibrium with 
the magnetar surface radiation~\citep{Cordes08} is below the iron evaporation point of about 2000~K at $R_b$. 
Therefore, our scenario favors the association of FRBs with relatively old and massive NS.  

Non-thermal radiation from the magnetar, i.e., the low-frequency Poynting flux-dominated flow close to the star 
and the flux of X-rays and gamma-rays associated with pair-creation farther out~\citep{Kirk09},
may also cause the evaporation of the asteroid.
For the low-frequency Poynting flux-dominated flow from the magnetar, since the ratio of the size of the body to the incident wavelength is
$r_{\rm a}/(c P) = 2 \times 10^{-5} \ll 1$, the corresponding evaporation should be negligible according to the Mie theory~\citep{Kotera16}.
Heating by high energy particles may evaporate an orbiting asteroid at a close distance from the magnetar.
However, since the asteroid is falling and 
the fraction of particles' kinetic energy in the pulsar's energy outflowing is decreasing when moving closer to the magnetar,
the asteroid would have not enough time to evaporate completely.
So heating by the non-thermal radiation from the magnetar also will not significantly affect our conclusion.

\subsection{X-ray Burst}

\begin{figure*}
	\begin{center}
		\includegraphics[scale=0.8]{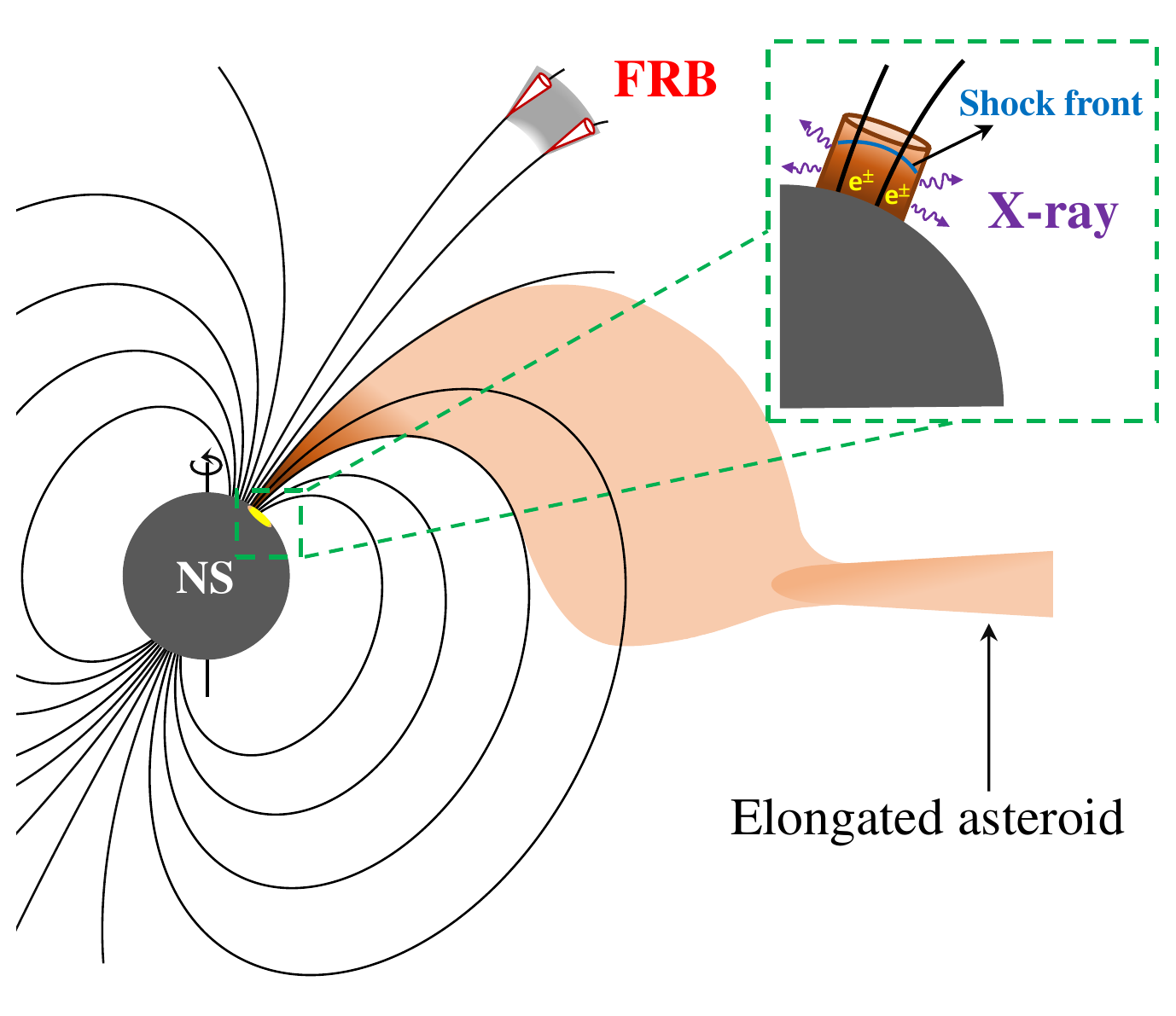}
		\caption{Schematic illustration of the impact between an asteroid and a magnetar.
		The asteroid gets elongated after entering $R_b$ and further accreted along field lines within $R_{\rm A}$.
		An instant accretion column is formed after collision onto the magnetar surface, 
		producing a Comptonized X-ray burst and an FRB in the magnetosphere.}
		\label{Fig1}
	\end{center}
\end{figure*}

Similar to the accretion of matter onto a compact object in binary systems,
an {\it instant accretion column} will form in our scenario~(see Figure~\ref{Fig1}).
As the accreting matter flows along magnetic field lines to the polar cap,
it will become highly supersonic and essentially in free-fall.
Before landing on the magnetar's surface, since the infalling material needs to be decelerated to subsonic,
some sort of strong shock must occur in the accretion stream.
After the shock, the hot gas settles to the NS surface in the magnetically confined column.
Since our understanding of the column structure/shock characteristics is far from complete,
here, we adopt a simple model to estimate basic parameters for the instant accretion column.  

The isotropic fluence of the X-ray burst is $\sim 7 \times 10^{-7}$~erg~cm$^{-2}$, lasting for $\delta t \sim 150$~ms.
We attribute the X-ray burst energy to the release of gravitational potential energy of the asteroid, i.e.,
\begin{equation}
E_{X} = \eta \frac{G M_{\rm NS} m_{\rm a}}{R_{\rm NS}},
\end{equation}
where $\eta$ is the efficiency of energy transforming.
Taking $E_{X} = 8 \times 10^{39}$~erg and typical $\eta \sim 0.1$,
the asteroid mass needed is $m_{\rm a} = 4 \times 10^{20}$~g.
This mass is roughly in the mass range of normal asteroids.
The average X-ray luminosity is $L_X = E_{X}/\delta t = 5 \times 10^{40}$~erg~s$^{-1}$,
and the corresponding accretion rate is $\dot{M} = m_{\rm a}/ \delta t= 2.7 \times 10^{21}$~g~s$^{-1}$.
The X-ray luminosity of the burst exceeds the Eddington luminosity ($L_{\rm Edd} \approx 2 \times 10^{38} M_{\mathrm{NS}, 1.4 M_{\odot}}$~erg~s$^{-1}$), 
indicating the column is radiatively dominated.
Due to the increasing magnetic pressure approaching the magnetar surface, 
at the magnetar's surface, we assume that the cross-section of the accretion column is a thin rectangle of width $d$ and length $l$.
The polar angle (respect to the magnetic axis) of this rectangle could be estimated
by the landing point of accreted material, i.e., $\theta_{\rm land} = \arcsin (\sqrt{R_{\rm NS}/R_{\rm A}})$.
Then $l$ could be estimated as $l \approx 2 \pi R_{\rm NS} \sin \theta_{\rm land} \times \delta t / P = 8 \times 10^4$~cm.
The free-fall velocity at the surface is $v_{\rm ff} = \sqrt{2 G M_{\rm NS} / R_{\rm NS}}$.
After the shock, the plasma falls down to a velocity of $v_{\rm fd} = v_{\rm ff}/7$~\citep{Lyubarskii82,Becker98}
and the mass density changes as $\rho_{\rm ps} = \dot{M} / (l d v_{\rm fd})$.

Following the calculations of \cite{Mushtukov15}, the height of the accretion column $H$ can be estimated from
\begin{equation}
L_{X} \approx 7.6 \left( \frac{l/d}{10} \right) \frac{\kappa_{\mathrm T}}{\kappa_{\perp}} f\left(\frac{H}{R_{\rm NS}}\right) L_{\rm Edd},
\end{equation}
where $f(\frac{H}{R_{\rm NS}}) = \ln(1+\frac{H}{R_{\rm NS}})-\frac{H}{R_{\rm NS}+H}$,
$\kappa_{\mathrm T}$ is the Thomson electron scattering opacity of the solar mix plasma (the mean number of nucleons per electron is 1.17),
and $\kappa_{\perp}$ is the opacity across the field lines.
For a strong magnetic field and a typical photon energy $E_{\gamma}$ ($\sim 84$~keV for this burst) less than cyclotron energy $E_{\rm cycl}$,
$\kappa_{\mathrm T}/\kappa_{\perp} \sim (E_{\gamma}/E_{\rm cycl})^{-2} \sim 1000$~\citep{Canuto71}.
We assume thermodynamical equilibrium deep inside the colunm and the plasma temperature to be $T$, 
the radiation field energy density in the deep center could be written as
\begin{equation}
P_{\rm rad} \approx \rho_{\rm ps} \frac{G M_{\rm NS}}{R_{\rm NS} + H} \frac{H}{R_{\rm NS}} \approx \frac{a T^4}{3},
\end{equation}
where $a$ is the radiation constant.
An empiric value of $l/d \sim 10-100$ is usually adopted in previous studies~\citep{Basko76,Mushtukov15}.
We find that our scenario could well explain the spectral feature of the X-rays 
if we chose $d = l/10 = 8 \times 10^3$~cm.
Taking $L_X = 5 \times 10^{40}$~erg~s$^{-1}$,
we obtain $H / R_{\rm NS} = 0.3$, $T \approx 2 \times 10^9$~K or $k T \approx 170$~keV ($k$ is the Boltzmann constant).
On the other hand, the effective temperature ($T_{\perp}$) corresponding to the escaping flux (perpendicular to the field lines) 
is determined by the emergent flux
\begin{equation}
F_{\perp,\rm esp} \approx \frac{4 c P_{\rm rad}}{\rho_{\mathrm{ps}} \kappa_{\perp} d} = \sigma_{\rm SB} T_{\perp}^4,
\end{equation}
where $\sigma_{\rm SB}$ is the Stefan-Boltzmann constant.
This relation gives $k T_{\perp} \approx 40$~keV, roughly consistent with
the cutoff energy $\sim 80$~keV of the X-ray burst~\citep{Li20}.

More detailed processes should be considered for the radiation from the spot.
High temperatures of the column result in the creation of electron-positron pairs,
so that soft X-ray photons will be bulk Comptonized by $e^+e^-$ pair plasmas before escaping. 
The pair number density could be given by \citep{Mushtukov19}
\begin{equation}
n_{\pm} \simeq 1.8 \times 10^{30} e^{-m_e c^2/kT} \left( \frac{k T}{m_e c^2} \right)^{3/2}~\mathrm{cm}^{-3},
\end{equation}
from which we get $n_{\pm} \simeq 3 \times 10^{28}$~cm$^{-3}$ using $k T = 170$~keV.
Considering the total energy of accreting material is going into pair creation, 
the Lorentz factor of the pairs $\gamma_{\pm}$ could be obtained from 
\begin{equation}
(\Gamma_{\rm ff} - \Gamma_{\rm fd}) \rho_{\mathrm{ps}} c^2 = 2 \gamma_{\pm} n_{-} m_e c^2,
\end{equation}
where $\Gamma_{\rm ff/fd} = (1-(v_{\rm ff/fd}/c)^2)^{-1/2}$, which gives
$\gamma_{\pm} \simeq 8$.
The scattering between the seed thermal/bremsstrahlung photons of $\sim 1$ keV and 
$e^+e^-$ pairs will result in hard photon energy of 60 keV~\citep{Becker07},
which is also consistent with the cutoff energy of this X-ray burst.
To obtain an accurate photon spectrum, we need to solve the Kompaneet's equation, which is beyond the scope of this work.
Scattering, recoil and escaping from a finite medium should be included.
However, according to studies of unsaturated Compton scattering in X-ray binaries, 
the emerging spectrum is expected to be a cut-off power law shape~\citep{Ferrigno16},
which is consistent with the spectral feature of the X-ray burst here.

The X-ray burst showed two hard peaks with a separation of $\sim 30$~ms.
This could be understood by supposing that the elongated asteroid breaks into two main fragments around $R_b$~\citep{Dai20}.
We can infer that the distance between these two fragments should be roughly $3.5 \times 10^6$~cm,
if attribute the time separation of X-ray peaks to difference of the arrival time of two fragments~(using Equation (3) in \citealt{Geng15}).
This distance is comparable with the radius of the asteroid before elongation, making the explanation more plausible.

\subsection{Coherent Radio Pulse}

FRB 200428 consists of two sub-bursts with durations of $\sim 0.6$ ms~and $\sim 0.34$~ms respectively,
separated by $\sim 30$~ms~\citep{CHIME20b}. 
For the first sub-burst, the primary beam detection from STARE2 gives that its isotropic-equivalent
luminosity is $L_{\rm FRB} = 4.2 \times 10^{38}$~erg~s$^{-1}$ at band $\sim 1.4$~GHz~\citep{Bochenek20}.
The second sub-burst is a bit fainter. Below, we will take the first sub-burst as the typical case in our interpretation.

It is still uncertain how coherent radio emission is produced from pulsars.
Coherent mechanisms, either by charged bunches (e.g., \citealt{Ruderman75,Benford77,Lyutikov16,Yang18,Kumar20}) 
or a maser mechanism (e.g., \citealt{Blandford75,Luo92,Lyubarsky14,Metzger19}) due to growth of plasma instabilities
are proposed to produce radio emission.
The same situation has also confronted us in FRB.
Since the time delay between the radio signal and the X-ray burst is in good agreement with 
the dispersion delay, both of them are suggested to originate within the magnetosphere,
which motivates us to explain the radio pulse with curvature radiation from bunches.

As mentioned in the previous subsection, the landing of accretion material onto the NS surface
produces abundant $e^+e^-$ pairs.
Considering that the landing point of the column $\theta_{\rm land}$ is beyond the gap region of the NS,
only a small portion of these pairs will flow into the gap region.
On the other hand, the large electric field in the gap can accelerate electrons and produce pair cascade within a very short
timescale $\sim 1$~$\mu$s~\citep{Mitra17}.
During the rising phase of X-ray burst, the arrival of the charges will screen the gap electric field
due to the continuous inreasing supply of charges.
The screen is destructed once the charge density from the accretion column decreases to a threshold
during the decline phase of the X-ray burst.
At that time, these charges from the column will be accelerated (with Lorentz factors of $\gamma$) to flow along the open field lines. 
After the leaving of this plasma cloud, the discharge from the gap can initiate soon within a period of $\sim 1 \mu$s,
which produces a following plasma cloud.
When it caught up with the plasma cloud emitted from accretion material, the overlapping of the slow and fast 
moving particles will lead to the two-stream instability in the plasma, hence the coherent emission.
The height of the gap is~\citep{Mitra17}
\begin{equation}
h = 95 B_{\rm NS,14}^{-4/7} P^{1/7} \dot{P}^{-2/7}_{-11} R_{\rm NS, 6}^{2/7}~\mathrm{cm}.
\end{equation}
The velocity difference between the slow and fast moving particles is about 
$\Delta v = c/(2\gamma^2)$ and the time for the particles to overlap is $h/\Delta v$.
Hence, the instability can develop at a distance of $r_{\rm emi} \sim c h/\Delta v \simeq 2 \gamma^2 h$ from NS.
The corresponding characteristic frequency of curvature emission is
\begin{equation}
\nu_c = \gamma^3 \frac{3 c}{4 \pi r_{\rm c}},
\end{equation}
where $r_{\rm c}$ is the curvature radius and is related to $r_{\rm emi}$ 
by $r_{\rm c} \simeq \frac{4 r_{\rm emi}}{3 \sin \theta_{\rm emi}}$ according to the magnetosphere geometry,
and $\theta_{\rm emi}$ is the poloidal angle of the emission region.
Therefore, we could obtain
\begin{equation}
\gamma = 677 \left(\frac{\nu_{\rm c}}{1.4~\mathrm{GHz}}\right) \left(\frac{h}{65~\mathrm{cm}}\right) \left(\frac{\sin \theta_{\mathrm{emi}}}{0.05}\right)^{-1}.
\end{equation}
This is witin the range of Lorentz factor ($10^2 - 10^4$) of $e^+e^-$ accelerated from the NS gap.

Assuming the length of the bunching shell is $\Delta$,
the duration $\delta t$ of the FRB pulse implies $\Delta \simeq c \delta t$.
We further assume that the ratio of the shell's solid angle to $4 \pi$ is $f$,
then the emission volume is $V_{\rm emi} = 4 \pi f r_{\rm emi}^2 \Delta$.
Electrons radiate coherently in patches with a characteristic radial size of $\lambda = c/\nu_c$,
and they can be casually connected in the relativistic beam of angle of $1/\gamma$,
the corresponding volume of one small patch is $V_{\rm coh} = (4/\gamma^2) r_{\rm emi}^2 \lambda$.
The number of the patches writes as $N_{\rm pat} \approx V_{\rm emi}/V_{\rm coh}$.
Then the coherent curvature emission luminosity can be estimated as
\begin{equation}
L_{\rm coh} = (P_e N_{\rm coh}^2) \times N_{\rm pat},
\end{equation}
where $N_{\rm coh} = n_e \times V_{\rm coh}$ is the number of electrons in each patch.
Using $L_{\rm coh} = f L_{\rm FRB}$, we can constrain the electron number density to be 
\begin{eqnarray}
n_e &\simeq& 4.6 \times 10^9 \left( \frac{L_{\rm FRB}}{4.2 \times 10^{38}~\mathrm{erg~s}^{-1}} \right)^{1/2} 
\left(\frac{\gamma}{677} \right)^{1/2} \\ \nonumber
& & \left(\frac{r_{\rm emi}}{6 \times 10^7~\mathrm{cm}} \right)^{-3/2} 
\left(\frac{\delta t}{0.6~\mathrm{ms}} \right)^{-1/2} \left(\frac{\sin \theta_{\mathrm{emi}}}{0.05} \right)^{-1/2}
\mathrm{cm}^{-3}.
\end{eqnarray}
The corresponding plasma density near the NS surface should be 
$\sim n_e (r_{\rm NS}/r_{\rm emi})^{-2} \simeq 1.7 \times 10^{13}$~cm$^{-3}$.
The \cite{Goldreich69} charge number density at the cap region is  
$n_{\rm GJ}(r_{NS},\theta_{\rm open}) =  3.8 \times 10^{10} \mathrm{cm}^{-3}$,
where $\sin \theta_{\rm open} = \sqrt{R_{\rm NS}/R_{\rm LC}}$.
Therefore, a reasonabale pair multiplicity factor of $\sim 400$ is needed during the spark.

In our scenario, an FRB is emitted along open field lines at $r_{\rm emi} \sim 60~R_{\rm NS}$,
which indicates that the edge of the lightning cone of this magnetar may sweep the line of sight.
Although the radiation position for the FRB and periodic radio pulsations may be different,
the detection of periodic radio pulsations may still be possible since we have 
little constraints on the $f$ factor of the FRB.
Interestingly, a $\sim$ 7-sigma detection of periodic radio pulsations from SGR 1935+2154 
was reported very recently \citep{Burgay20}. 
If such detection is confirmed in the future, our scenario would be highly favored.

Usually, the direct collision between asteroids and the magnetar may be very rare
since recurrence time of such strong direct impacts ($m_{a} \ge 10^{18}$~g) in a typical NS planetary system 
can be as large as $\sim 10^6 - 10^7$~years according to previous studies~\citep{Tremaine86,Litwin01}. 
However, in some special cases, it may still be possible that the NS is passing through an asteroid belt
as suggested by \cite{Dai16} so that several FRBs could be produced within a short period due to multiple collisions.
A much weaker (30 mJy) radio pulse has been observed by Five-hundred-metre Aperture Spherical radio telescope (FAST) 
two days after the two FRB pulses~\citep{Zhang20}.
It is doubtful if this weak burst can be regarded as an FRB or not because its
equivalent isotropic energy is almost $10^8$ times smaller than the two pulses.
If it were really from SGR 1935+2154,
it may be caused by pollution of NS magnetospheres by low-level accretion from a circumpulsar debris disk~\citep{Cordes08}.

\subsection{Follow-up Burst}

The landing of asteroid material onto the magnetar will increase the moment of inertia of the magnetar,
leading to an abrupt change of rotational frequency of magnetar~\citep{Huang14}.
Assuming the moment of inertia of the magnetar crust $I_c$ is only one percent of
the total moment of inertia, i.e., $I_c = 0.01 \times \frac{2}{5} M_{\rm NS} R_{\rm NS}^2 \sim 10^{43}$~g~cm$^2$.
The asteroid of mass $4 \times 10^{20}$~g will lead to an instantaneous change in frequency
$\Delta \nu \sim -1.0 \times 10^{-12}$ Hz, which is too small to be detected from X-ray/radio observation.
However, we have ignored the orbital angular momentum of the asteroid itself in the above estimate.
The transfer of asteroid's orbital angular momentum to the magnetar crust could
result in a $\Delta \nu$ of magnitude $\sim 10^{-8}$ Hz.
Future observations of magnetars with radio pulses could help to identify this hypothesis.

The impact of the accretion column against the magnetar surface may also
lead the crust to be more unstable, i.e., trigger the crustquakes~\citep{Thompson95}
or magnetic field line reconnection~\citep{Lyutikov03}. 
This process heats the surface in one or more regions, from which seed thermal photons
might be resonant cyclotron scattered by the electrons.
As a result, an X-ray burst forest is expected to follow the FRB-associated X-ray burst.
Since these following X-ray bursts come from the normal activities of an isolated magnetar, 
they are unlikely to be associated with further FRBs.

\section{Summary and Discussion}

We study the impact between an asteroid and a magnetar,
trying to explain the association of the Galaxy FRB 200428 with an X-ray burst.
For an asteroid with a mass of $\sim 10^{20}$~g falling onto the magnetar's surface, 
an instant accretion column will be formed, which produces a Comptonized X-ray burst
and an FRB in the magnetosphere. 
Our calculations show that all the observational features could be self-consistently
explained within this scenario.   

Both the radio and the X-ray burst showed two peaks with a separation of $\sim 30$~ms.
Similar substructures also exist in some extragalactic FRBs. 
This may be due to the unsmooth disruption process of the asteroid.
At $R_b$, the asteroid may break into two or even more fragments~\citep{Dai20}, making the later accretion
flow and emission to be intermittent.

The accretion column in our scenario may be identified indirectly from the temporal variability of the burst.
It is known that the postshock flow may be subject to a global thermal instability
revealed by rapid variations of the cooling time scale with the shock speed.
The instability drives the shock front to oscillate with respect to its stationary position, 
causing fluctuations in the emission flux~\citep{Saxton98,Mignone05}. This phenomenon has been detected in
accreting white dwarfs, which show optical emission with quasi-periodic oscillations (QPOs) 
$\sim$ several Hz~\citep{Imamura91,Mouchet17,Bera18}.
Unfortunately, radiative shocks in the accretion column onto magnetars are poorly studied
in the literature, especially including the photon-$e^+e^-$ pairs scattering processes. 
It is difficult for us to estimate the oscillating frequency within our current work.
Nevertheless, the QPOs (if exist) in the temporal variability of this X-ray burst from SGR 1935+2154 will strongly favor our scenario.
At the same time, the QPO should emerge only in the burst associated with the FRB 200428,
rather than the following bursts.

By now, the consensus on FRBs is that the radio emission of FRBs should be coherent.
On the other hand, we are confident that coherent radio emission comes from the 
moving plasma cloud in the magnetosphere according to observations of pulsars.
Therefore, what we have proposed belongs to processes that can trigger the coherent emission from the magnetosphere of NSs,
resulting in a millisecond burst rather than normal pulsar pulses.
It should be noted that multiple NS-asteroid impacts may produce repeating FRBs/periodic FRBs (see \citealt{Dai16,Dai20b} for details).
Increasing observational constraints will help us to judge whether repeating FRBs have a unique or various origins in the future~\citep{Smallwood19}.

\acknowledgments

We thank the anonymous referee for valuable suggestions.
We also thank Shuang-Nan Zhang, Andr\'e Oliva and Zi-Gao Dai for helpful discussion. 
This work is partially supported by the National Natural Science Foundation of China
(grants No. 11903019, 11873030, U1938201 and U1838113),
and by the Strategic Priority Research Program of the Chinese Academy of Sciences 
(``multi-waveband Gravitational Wave Universe'' grant No. XDB23040000,
XDA15360300 and XDA15052700).
RK acknowledges financial support via the Emmy Noether Research Group on Accretion Flows 
and Feedback in Realistic Models of Massive Star Formation funded by 
the German Research Foundation (DFG) under grant no. KU 2849/3-1 and KU 2849/3-2.
LBL acknowledges support from the Natural Science Foundation of Hebei Province of China 
(grant No. A2020402010).

\end{document}